\documentstyle[epsfig,12pt]{article}  
\newcommand{\bce}{\begin{center}}
\newcommand{\ece}{\end{center}}
\newcommand{\beq}{\begin{equation}}
\newcommand{\eeq}{\end{equation}}
\newcommand{\bea}{\vspace{0.25cm}\begin{eqnarray}}
\newcommand{\eea}{\end{eqnarray}}

\newcommand{\ba}{\begin{array}}
\newcommand{\ea}{\end{array}}

\newcommand{\r}{\mbox{{\boldmath
$\rho$}}}
\newcommand{\ta}{\mbox{{\boldmath
$\tau$}}}

\newcommand{\qb}{\mbox{{\bf
q}}}

\newcommand{\rb}{\mbox{{\bf
r}}}

\def\lsim{\mathrel{\rlap{\lower4pt\hbox{\hskip1pt$\sim$}}
    \raise1pt\hbox{$<$}}}         %less than or approx. symbol
\def\gsim{\mathrel{\rlap{\lower4pt\hbox{\hskip1pt$\sim$}}
    \raise1pt\hbox{$>$}}}         %greater than or approx. symbol

    \def\beq{\begin{equation}}
    \def\endeq{\end{equation}}
    \def\bea{\begin{eqnarray}}
    \def\arr{\begin{eqnarray}}
    \def\eea{\end{eqnarray}}

\def\q2{$Q^{2}$}
\def\s2{2$S$}

\begin{document}
%\doublespace
\thispagestyle{empty}
\vspace*{-2cm}
\begin{flushleft}  
\hspace{7.3cm}Talk given at 34th Rencontres de Moriond\\
\hspace{7.3cm}"QCD and High Energy Hadronic Interactions",\\
\hspace{7.3cm}Les Arcs, France, March 20--27, 1999\\
\end{flushleft}
\bigskip
\vspace{2cm}

\begin{center}

  {\large\bf
TRANSVERSE SPECTRA OF INDUCED RADIATION
\\
\vspace{1.5cm}
  }
\medskip
  {\large
  B.G. Zakharov
  \bigskip
  \\
  }
  {\it
Landau Institute for Theoretical
Physics,
  GSP-1, 117940,\\  Kosygina Str. 2, 117334 Moscow,
  Russia
  \vspace{2.7cm}\\
  }
  
  {\bf
  Abstract}
\end{center}
{
\baselineskip=9pt

Transverse spectra of induced radiation are discussed within 
the light-cone path integral approach to the LPM effect. The
results are applicable in both QED and QCD.
}
\pagebreak
%-------------------------------------------------------------
\newpage
%-------------------------------------------------------------

Recently the Landau-Pomeranchuk-Migdal (LPM) effect \cite{LP,Migdal} 
in induced radiation in QED and QCD
has attracted much attention (see review by Klein \cite{Klein} and 
references therein).
Understanding the LPM effect in QCD is of great importance for
evaluation of parton 
energy loss in nuclei and a hot 
QCD medium \cite{BDPS,BGZ1,BDMPS1,BGZ2,BDMS}.
The case of hot QCD medium is especially 
interesting in view of the experiments on $AA$-collisions
at RHIC and LHC. 

In Ref. \cite{BGZ1} I have developed a new rigorous light-cone 
path integral approach 
to the LPM effect. There I have discussed
the $p_{\perp}$-integrated spectra. In this talk I discuss 
the transverse spectra of induced 
radiation. Similarly to Ref. \cite{BGZ1} the results are applicable 
in both QED and QCD. For simplicity I describe the formalism  
for an induced $a\rightarrow bc$ transition in QED for scalar particles
with an interaction Lagrangian  
$L_{int}=\lambda [\hat\psi_{b}^{+}\hat\psi_{c}^{+}\hat\psi_{a}+
\hat\psi_{b}\hat\psi_{c}\hat\psi_{a}^{+}]$.
The corresponding $S$-matrix element reads
\beq
\langle bc|\hat{S}|a\rangle=i\int\! dt
d\rb \lambda
\psi_{b}^{*}(t,\rb)\psi_{c}^{*}(t,\rb)\psi_{a}(t,\rb)\,,
\label{eq:1}
\eeq
where $\psi_{i}$ are the wavefunctions (ingoing for $i=a$ 
and outgoing for $i=b,c$). I normalize the flux to unity
at $z=-\infty$ for $i=a$ and at $z=\infty$ for $i=b,c$, and 
write $\psi_{i}$ as
\beq
\psi_{i}(t,\rb)=\frac{1}{\sqrt{2E_{i}}}\exp[-i(t-z)p_{i,z}]
\phi_{i}(t,\rb)\,.
\label{eq:2}
\eeq
In the high energy limit, $E_{i}\gg m_{i}$,
the dependence of $\phi_{i}$ on the variable 
$\tau=(t+z)/2$ at $t-z=$const is governed by the two-dimensional 
Schr\"odinger equation
\beq
i\frac{\partial{\phi_{i}}}{\partial{\tau}}=
H_{i}\phi_{i}\,,
\label{eq:3}
\eeq
\beq
H_{i}=-\frac{\Delta_{\perp}}{2\mu_{i}}+e_{i}A^{0}+\frac{m_{i}^{2}}{2\mu_{i}}\,,
\label{eq:4}
\eeq
where $\mu_{i}=p_{i,z}$, $e_{i}$ is the electric charge, $A^{0}$ 
is the potential of the target.

After some algebra from (1), (2) one can obtain in the high energy 
limit the following
expression for the inclusive probability of induced radiation 
\beq
\frac{d^{5}P}{dx d\qb_{b}d\qb_{c}}=\frac{2}{(2\pi)^{4}}
\mbox{Re}\!
%\frac{g}{(2\pi)^{4}}
\int
\limits_{z_{1}<z_{2}}\!
d\r_{1}d\r_{2}dz_{1}dz_{2}\,
g\langle F(z_{1},\r_{1})
F^{*}(z_{2},\r_{2})\rangle\,,
\label{eq:5}
\eeq
where 
$\qb_{b,c}$ are the transverse momenta, $\r$ is transverse coordinate,
$x=p_{b,z}/p_{a,z}$ (note that for the particle $c$  
$p_{c,z}=(1-x)p_{a,z}$),
$g=\lambda^{2}/[16\pi x(1-x)E_{a}^{2}]$,
$\langle ...\rangle $
means averaging over the states of the target,
%
%$F(z,\r)=\phi_{b}^{*}(t=z,\rb)
%\phi_{c}^{*}(t=z,\rb)\phi_{a}(t=z,\rb)$.
$\left.F(z,\r)=\phi_{b}^{*}(t,\rb)
\phi_{c}^{*}(t,\rb)\phi_{a}(t,\rb)\right|_{t=z}$.
Since the wavefunctions enter (\ref{eq:5}) only at $t=z$,
$\phi_{i}$ can be regarded 
as functions of $z$, and $\r$. 
In the Schr\"odinger equation (\ref{eq:3})
$z$ will play the role of time.
I represent $z$-dependence of $\phi_{i}$ in terms of the 
Green's function, $K_{i}$, of the Hamiltonian (\ref{eq:4}).
Then, in the diagram language (\ref{eq:5})
is described by the graph of Fig.\,1a. I depict 
$K_{i}$ ($K_{i}^{*}$) by $\rightarrow$ ($\leftarrow$).
The dotted line
shows the transverse density matrices at large longitudinal 
distances
in front of ($z=z_{i}$) and behind ($z=z_{f}$) 
the target.\footnote{Strictly speaking, in 
(\ref{eq:1}), (\ref{eq:5})
the adiabatically vanishing at $|z|\sim |z_{i,f}|$ coupling 
should be used. 
For simplicity I do not indicate
the coordinate dependence of the coupling.}
If the particle $a$ is produced in a hard reaction, and 
does not propagate from infinity,
then $z_{i}$ equals the coordinate 
of the production point. 
\begin{figure}[h]
\begin{center}
\psfig{file=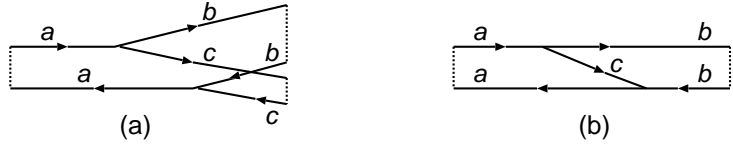,height=2cm}
\end{center}
\vspace{-.3cm}
\caption[.]{The diagram representation of the inclusive spectrum 
(5) (a), and (6) (b).}
\end{figure}
Below I will consider the radiation rate
integrated over $\qb_{c}$. In this case the graph of Fig.\,1a is
transformed into the one of Fig.\,1b. The corresponding analytical
expression reads
\bea
\frac{d^{3}P}{dx d\qb_{b}}=\frac{2}{(2\pi)^{2}}
\mbox{Re}
%\frac{g}{(2\pi)^{2}}
\int
\limits_{z_{i}}^{z_{f}}\!dz_{1}\int\limits_{z_{1}}^{z_{f}}\!dz_{2}
\int\!
d\r_{b,f}d\r_{b,f}^{'}d\r_{b}d\r_{b}^{'}
d\r_{a}d\r_{a}^{'}d\r_{a,i}d\r_{a,i}^{'}
\,g\exp[-i\qb_{b}(\r_{b,f}-\r_{b,f}^{'})]\nonumber \\
\times S_{b}(\r_{b,f},\r_{b,f}^{'},z_{f}|
\r_{b},\r_{b}^{'},z_{2})
M(\r_{b},\r_{b}^{'},z_{2}|
\r_{a},\r_{a}^{'},z_{1})
S_{a}(\r_{a},\r_{a}^{'},z_{2}|
\r_{a,i},\r_{a,i}^{'},z_{i})\,,
\,\,\,\,\,\,\,\,\,\,\,\,\,\,\,\,
\label{eq:6}
\eea
%where the factors
%$S_{a,b}$ and $M$ are given by
\beq
S_{i}(\r_{2},\r_{2}^{'},z_{2}|
\r_{1},\r_{1}^{'},z_{1})=
\langle 
K_{i}(\r_{2},z_{2}|\r_{1},z_{1})
K_{i}^{*}(\r_{2}^{'},z_{2}|\r_{1}^{'},z_{1})
\rangle\,,
\label{eq:7}
\eeq
\beq
M(\r_{2},\r_{2}^{'},z_{2}|
\r_{1},\r_{1}^{'},z_{1})=
\langle 
K_{b}(\r_{2},z_{2}|\r_{1},z_{1})
K_{c}(\r_{2}^{'},z_{2}|\r_{1},z_{1})
K_{a}^{*}(\r_{2}^{'},z_{2}|\r_{1}^{'},z_{1})
\rangle\,.
\label{eq:8}
\eeq
Using the path integral representation for the Green's
functions one can evaluate analytically the initial- and
final-state interaction factors $S_{a,b}$ \cite{BGZ3}. 
The factor $M$ (\ref{eq:8}) differs from that of Ref. \cite{BGZ1}
by the replacement of $\r_{2}^{'}$ by $\r_{2}$ in the 
Green's function $K_{c}$. Similar to that of Ref. \cite{BGZ1} it can be
expressed through the Green's function $K_{bc}$ describing
the relative motion of the particles $b$ and $c$
in a fictitious $\bar{a}bc$ system. After analytical 
integration over the center-of-mass transverse coordinates
the radiation rate takes the form
\bea
\frac{d^{3}P}{dx d\qb_{b}}=\frac{2}{(2\pi)^{2}}
\mbox{Re}
%\frac{g}{(2\pi)^{2}}
\int
\limits_{z_{i}}^{z_{f}}\!dz_{1}\int\limits_{z_{1}}^{z_{f}}\!dz_{2}
\int\!
d\ta_{b}\,g\exp(-i\qb_{b}\ta_{b})
\nonumber\\
\times
\Phi_{b}(\ta_{b},z_{2})
\exp\left[\frac{i(z_{1}-z_{2})}{L_{f}}\right]
K_{bc}(\ta_{b},z_{2}|0,z_{1})
\Phi_{a}(\ta_{a},z_{1})
\,,
\label{eq:9}
\eea
where
\beq
\Phi_{a}(\ta_{a},z_{1})=
\exp\left[-\frac{\sigma_{a\bar{a}}(\ta_{a})}{2}
\int\limits_{z_{i}}^{z_{1}}\!dz n(z)\right]\,,
%\label{eq:10}
%\eeq
%\beq
\,\,\,\,\,\,\,\,
\Phi_{b}(\ta_{b},z_{2})=
\exp\left[-\frac{\sigma_{b\bar{b}}(\ta_{b})}{2}
\int\limits_{z_{2}}^{z_{f}}\!dz n(z)\right]
\label{eq:11}
\eeq
are the eikonal initial- and final-state absorption
factors,\footnote{
I emphasize, that appearance of the eikonal absorption
factors in (\ref{eq:9}) is a nontrivial consequence 
of the specific form of evolution operators
$S_{a,b}$ \cite{BGZ3}, and is not connected with applicability
of the eikonal approximation.}
$\ta_{a}=x\ta_{b}$,
$L_{f}=2E_{a}x(1-x)/[m_{b}^{2}(1-x)+m_{c}^{2}x-m_{a}^{2}x(1-x)]$.
The Hamiltonian for the Green's function $K_{bc}$ reads
\beq
H_{bc}=-\frac{\Delta_{\perp}}{2\mu_{bc}}-
\frac{in(z)\sigma_{\bar{a}bc}(\ta_{bc},\ta_{ab})}{2}\,,
\label{eq:12}
\eeq
where $\mu_{bc}=E_{a}x(1-x)$, 
$\ta_{ab}=-[\ta_{a}+\ta_{bc}(1-x)]$.
In (\ref{eq:11}), (\ref{eq:12})
$n(z)$ is the number density of the target, $\sigma_{a\bar{a}}$
and $\sigma_{b\bar{b}}$ are the dipole cross sections of
interaction with the medium constituent of $a\bar{a}$ and
$b\bar{b}$ pairs, and $\sigma_{\bar{a}bc}$ is the three-body 
cross section for $\bar{a}bc$-system. 

%Equation (\ref{eq:9}) generalizes our result \cite{BGZ1} 
%for $p_{T}$-integrated spectra, which can be obtained from
%(\ref{eq:9}) after integration over $\qb_{b}$.
%Note, that contrary to the case of
%$p_{T}$-integrated cross section, the potential in (\ref{eq:12})
%depends on the azimuthal angle.

The integration over $\qb_{b}$ in (\ref{eq:9})
gives the $x$-spectrum
\bea
\frac{dP}{dx}=2\mbox{Re}
\int
\limits_{z_{i}}^{z_{f}}\!dz_{1}\int\limits_{z_{1}}^{z_{f}}\!dz_{2}
\,g\exp\left[\frac{i(z_{1}-z_{2})}{L_{f}}\right]
K_{bc}(0,z_{2}|0,z_{1})
\,.
\label{eq:9p}
\eea
In Ref. \cite{BGZ1} I have derived the $p_{T}$-integrated 
radiation rate using
the unitarity connection between 
the probability of $a\rightarrow bc$ transition 
and the radiative correction to $a\rightarrow a$ transition. 
The latter is described by the diagram of Fig.\,2a, which 
can be transformed into the graph of Fig.\,2b, corresponding to
the integral in (\ref{eq:9p}).
\footnote{
The graph of Fig.\,2b (and the integral in (\ref{eq:9p}))
in itself 
requires subtracting of the infinite vacuum counter term.
The vacuum term has an imaginary part
connected with correction to $m_{a}$, which 
$\propto (z_{f}-z_{i})$, and a real part related to
the wavefunction renormalization. The latter appears
after separating the mass term, and connected with the  
configurations $z_{1}<z_{f}<z_{2}$. This boundary effect
is absent if the coupling vanishes at large $|z|$. 
Evidently, in this case the vacuum term does not affect the $x$-spectrum.
Nonetheless, it is convenient, as was done in Ref. \cite{BGZ1},
to keep the vacuum term to simplify the singular $z$-integration
in (\ref{eq:9p}).}
One can easy show that the diagram of Fig.\,2b can also be
obtained directly from that of Fig.\,1b after integration
over $\qb_{b}$.
\begin{figure}[h]
\begin{center}
\psfig{file=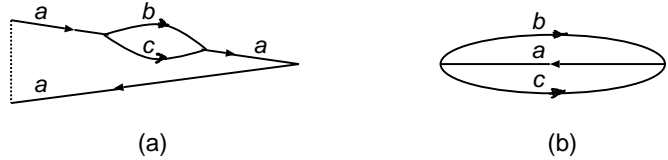,height=3cm}
\end{center}
\vspace{-.3cm}
\caption[.]{The diagram representation of the radiative correction 
to the probability of $a\rightarrow a$ transition.}
\end{figure}

Equation (\ref{eq:9}) establishes the theoretical basis for
evaluation of the $p_{T}$-dependence of the LPM effect.
Note that for transition with the formation length 
much greater than the target thickness (for the particle 
$a$ incident from infinity)
(\ref{eq:9}) can be
expressed through the light-cone wave function $\Psi_{a}^{bc}$ as
\bea
\frac{d^{3}P}{dx d\qb_{b}}=
\frac{2}{(2\pi)^{2}}\int\!
d\ta d\ta^{'} \exp(-i\qb_{b}\ta^{'})
\Psi_{a}^{bc*}(x,\ta-\ta^{'})\Gamma_{\bar{a}bc}(\ta_{bc},\ta_{ab})
\Psi_{a}^{bc}(x,\ta)
\,.
\label{eq:13}
\eea
Here,
$\ta_{bc}=\ta$, $\ta_{ab}=-[\ta (1-x)+\ta^{'}x]$,
%\beq
$
\Gamma_{\bar{a}bc}=
\left\{1-\exp\left[-\frac{\sigma_{\bar{a}bc}}{2}\!\int\! dz n(z)\right]
\right\}
%\label{eq:14}
%\eeq
$
is the Glauber profile function for interaction of $\bar{a}bc$
state with the target. The derivation of (\ref{eq:13}) is
based on a connection between the Green's function 
$K_{bc}$ in vacuum and $\Psi_{a}^{bc}$ \cite{SLAC1}.
Equation (\ref{eq:13})
generalizes the formula
for the $p_{T}$-integrated spectrum of Ref. \cite{NPZ}. It is of 
interest in its own right. In particular, the leading term in
$n(z)$ of the rhs in (\ref{eq:13}) gives a convenient
formula for evaluation of the Bethe-Heitler cross section 
through the light-cone wavefunction.

In general case one can estimate the inclusive
cross section using the parametrization
%\beq
$
\sigma_{\bar{a}bc}=C_{ab}\ta_{ab}^{2}+
C_{bc}\ta_{bc}^{2}+C_{ca}\ta_{ca}^{2}
%\label{eq:15}
%\eeq
$
(here $\ta_{ca}=-(\ta_{ab}+\ta_{bc})$).
Then the Hamiltonian (\ref{eq:12}) takes the oscillator
form with the frequency
$
\Omega(z)=\frac{(1-i)}{\sqrt{2}}
\left[\frac{n(z)C(x)}{E_{a}x(1-x)}\right]^{1/2}\,,
%\label{eq:16}
$
with 
$C(x)=C_{ab}(1-x)^{2}+
C_{bc}+C_{ca}x^{2}$.
The Green's function for the 
oscillator Hamiltonian 
%with the time-dependent frequency
can be written in the form
\beq
K_{osc}(\ta_{2},z_{2}|\ta_{1},z_{1})=
\frac{\gamma(z_{1},z_{2})}{2\pi i}
\exp\left\{i\left[\alpha(z_{1},z_{2})\ta_{2}^{2}
+\beta(z_{1},z_{2})\ta_{1}^{2}-\gamma(z_{1},z_{2})\ta_{1}\ta_{2}
\right]\right\}
\,,
\label{eq:17}
\eeq
where the functions $\alpha$, $\beta$ and $\gamma$ can be
evaluated in the approach of Ref. \cite{Jpsi}.
Using the parametrization $\sigma_{\bar{i}i}=C_{ii}\ta_{i}^{2}$
one can obtain 
\bea
\frac{d^{3}P}{dx d\qb_{b}}=\frac{1}{(2\pi)^{2}}\mbox{Re}
%\frac{g}{(2\pi)^{2}}
\int\limits_{z_{i}}^{z_{f}}
\!dz_{1}\int\limits_{z_{1}}^{z_{f}} \!dz_{2}\,g\,
\frac{\gamma(z_{1},z_{2})}{Q(z_{1},z_{2})}
\exp\left[-\frac{i\qb^{2}_{b}}{4Q(z_{1},z_{2})}
+\frac{i(z_{1}-z_{2})}{L_{f}}\right]
\,,
\label{eq:18}
\eea
where the factor $Q(z_{1},z_{2})$ can be expressed
through the parameters $C_{ij}$, the functions $\alpha$,
$\beta$, $\gamma$, $n$,  and $\Omega$. The formula 
for this factor is too cumbersome to be presented here.

The generalization of the above results to the realistic QED and QCD
Lagrangians reduces to a trivial replacement of the two- and 
three-body cross sections, and the vertex factor $g$. The latter, 
due to spin effects
in the vertex $a\rightarrow bc$, 
becomes an operator. The corresponding formulas 
are given in Refs. \cite{BGZ1,YAF}.
The results obtained can be applied to many 
problems. In particular in QCD this approach can be used for evaluation of
high-$p_{T}$ hadron spectra, the $p_{T}$-dependence
of DY pairs and heavy quarks production in $hA$-collisions, 
angular dependence of the parton energy
loss in hot QCD matter produced in $AA$-collisions. It is also
of interest for study the initial condition 
for quark-gluon plasma in $AA$-collisions.\\

%\section*{Acknowledgments}
I would like to thank N.N. Nikolaev and D. Schiff for
discussions. I am grateful to J.~Speth for the hospitality
at FZJ, J\"ulich, where this work was completed.   
This work was partially supported by the INTAS
grant 96-0597.

%\pagebreak
%\section*{References}

\end{document}